\documentclass[sn-basic]{sn-jnl}

\usepackage{graphicx}%
\usepackage{multirow}%
\usepackage{amsmath,amssymb,amsfonts}%
\usepackage{amsthm}%
\usepackage{mathrsfs}%
\usepackage[title]{appendix}%
\usepackage{xcolor}%
\usepackage{textcomp}%
\usepackage{manyfoot}%
\usepackage{booktabs}%
\usepackage{algorithm}%
\usepackage{algorithmicx}%
\usepackage{algpseudocode}%
\usepackage{listings}%

\newcommand{\ksmpc}{\,km\,s$^{-1}$\,Mpc$^{-1}$}    
\newcommand{\lcdm}{$\Lambda CDM$}
\newcommand{\wa}{$(w_0,w_a)$}
\newcommand{\ta}{$T_{\alpha}$}

\begin{document}

\title[Dark energy constraints]{Dark energy constraints from Pantheon+ Ia supernovae data}


\author*[1]{\fnm{Sergio} \sur{Torres-Arzayus}}\email{sergio.torres@icranet.org}
\affil*[1]{\orgname{International Center for Relativistic Astrophysics Network}, \orgaddress{\street{Piazza della Repubblica 10}, \city{Pescara}, \postcode{1-65122},  \country{Italy}}}



\abstract{
Measurements of the current expansion rate of the Universe, $H_0$, using standard candles, disagree with those derived from observations of the Cosmic Microwave Background (CMB). 
This discrepancy, known as the \emph{Hubble tension}, is substantial and suggests the possibility of revisions to the standard cosmological model (Cosmological constant $\Lambda$ and cold dark matter - \lcdm). 
Dynamic dark energy (DE) models that introduce deviations in the expansion history relative to \lcdm\ could potentially explain this tension. 
We used Type Ia supernovae (SNe) data to test a dynamic DE model consisting of an equation of state that varies linearly with the cosmological scale factor $a$. 
To evaluate this model, we developed a new statistic (the \ta\ statistic) used in conjunction with an optimization code that minimizes its value to obtain model parameters. The \ta\ statistic reduces bias errors (in comparison to the $\chi^2$ statistic) because 
it retains the sign of the residuals, which is meaningful in testing the dynamic DE model as the deviations in the expansion history introduced by this model act asymmetrically in redshift space. The DE model fits the SNe data reasonably well, but the available SNe data lacks the statistical power to discriminate
between \lcdm\ and alternative models. To further assess the model using CMB data, we computed the distance to the last scattering surface and 
compared the results with that derived from the \emph{Planck} observations. 
Although the simple dynamic DE model tested does not completely resolve the tension,
it is not ruled out by the data and could still play a role alongside other physical effects.
}



\maketitle

\section{Introduction}
\label{sec:intro}
\footnote{This version of the article has been accepted for publication, after peer
review but is not the Version of Record and does not reflect post-acceptance
improvements, or any corrections. The Version of Record is available online at:
https://doi.org/10.1007/s10509-024-04282-x.
Use of this Accepted Version is subject to the publisher’s Accepted
Manuscript terms of use https://www.springernature.com/gp/open-research/policies/accepted-manuscript-terms. }
The \emph{Hubble tension} consists of a $5\sigma$ discrepancy between the local value of the Hubble constant, $H_0$,
measured using \emph{standard candles} such as Type Ia Supernovae (SNe), and the value of $H_0$ derived from observations of the cosmic microwave background (CMB), assuming a flat-geometry cosmological model with a
cosmological constant $\Lambda$ 
and cold dark matter (\lcdm) Universe \citep{DiValentino}. In this paper we explore 
dynamic dark energy (DE) models that could explain the Hubble tension discrepancy.

The dominant methods for estimating $H_0$ from observations rely on \emph{standard candles} or \emph{standard rulers}
that are necessary for measuring distances.
The accuracy with which $H_0$ can be measured is limited by the precision in determining distances.
For sources at high redshifts, the distance determination is model dependent.
In the standard cosmological model, based on General Relativity and the Friedmann-Lema\^{i}tre-Robertson-Walker (FLRW) metric, 
the expansion rate of the universe is quantified by the Hubble parameter, $H(z) = \dot{a}/a$, 
where $z$ is the redshift, and $a$ is the cosmological scale factor,
$a = 1/(1 + z)$. 
The Hubble constant, $H_0$, is the value of the Hubble parameter, $H(z)$, at the present time, $z = 0$. 
It is important to note that although the estimation of the $H_0$ constant refers to $z = 0$, it is based on observations of sources at different $z$. 

The physics and methods used to derive a value for $H_0$ from the opposing sides of the Hubble tension
are fundamentally different, leaving ample room for disagreement. 
The local $H_0$ value is determined based on the relationship between magnitude (brightness) and redshift of standard candles, such as Type Ia SNe. Using this approach,
\citep[hereafter R22]{Riess2022} report a value of $73.04 \pm 1$ \ksmpc. 
On the other hand, the $H_0$ value derived from CMB is not a direct measurement, 
but is instead derived from fitting of a multi-parameter model to the observed CMB angular power spectrum, 
making it model dependent. 
The Planck Collaboration \citep[hereafter Planck-2018]{Planck2018} reports a value of $67.4 \pm 0.5$, assuming a \lcdm\ model. 
For simplicity, the units of $H_0$ are omitted hereafter (assume \ksmpc). 

To compute distances, R22 employs \emph{distance ladder} photometry calibration, 
in which Cepheid 
variable stars found in Type Ia SN host galaxies establish the connection between distance and brightness. 
Cepheid distances are determined from absolute magnitudes calibrated via the 
period-luminosity (P-L) relation (Leavitt Law). Systematic errors in Cepheid P-L calibration could  
potentially introduce errors in the local Hubble constant. However, recent high accuracy Cepheid photometry observations by the 
\emph{James Webb Space Telescope (JWST)} \citep{riess2023} 
conclude that errors attributable to Cepheid calibration do not significantly
contribute to the Hubble tension.
The Type Ia SN absolute magnitude, $M_B$, reported in R22 is $M_B = -19.253 \pm 0.027$ mag. If the Hubble tension discrepancy
were to be explained by a calibration error in SN Ia luminosity, the value of $M_B$ would need to be shifted by a delta of $-0.216$ mag, a scenario entirely ruled out by the \emph{JWST} results.

In addition to the SNe distance-redshift method there are various other
methods for estimating $H_0$, including, baryon acoustic oscillations (BAO) \citep{Ross_2015}, 
tip of the red giant branch (TRGB) \citep{Freedman_2019}, 
strong lensing time delays \citep{Suyu_2017}, 
and gravitational waves \citep{LIGO}. 
BAO and CMB approaches rely on statistical standard rulers, specifically the statistical galaxy-clustering scale (BAO) and the characteristic angular size of CMB fluctuations. 
While local measurements of $H_0$ tend to lean towards the higher end of the tension ($\sim 73$), high-redshift, 
BAO and early universe probes (CMB) tend towards the lower side of the tension ($\sim 67$).
Contrary to this trend, \citet{Tully_2023} derived a value of the Hubble constant of $76.9^{+8.2}_{-4.8}$ from an observation of a complete BAO structure, introducing complexity to the tension. 

One possible explanation for 
the Hubble tension is the occurrence of deviations in the acceleration of cosmic expansion compared to
what is predicted by the standard cosmological constant.
\citet{Dainotti_2021}, for instance, show that evolution of the Hubble parameter $H(z)$ away from standard \lcdm\ could reduce the $H_0$ tension.
Variations in the expansion could be triggered by dynamic 
dark energy (DE) equation of state (EOS) models. 
The EOS is defined as the ratio $w = P/\rho$, where $P$ and $\rho$ represent the DE pressure and density respectively, 
both in energy equivalent units to ensure a dimensionless ratio.  
In the standard \lcdm\ cosmology, DE density and pressure are constant, with $w = -1$.
However, if the EOS has undergone changes in the past, with $w(z)$ as a function of z, 
or equivalently as a function of the cosmological scale factor $a$,
then the expansion history could potentially account for the SN data that yields higher values of $H_0$,
while preserving the expansion behavior
at early times, where CMB observations are sensitive. 

Studying the origin of the acceleration of cosmic expansion has been
a significant endeavor since its initial discovery \citep{perlmutter1999}.
A plethora of models have been proposed, 
including (see \citet{DiValentino}), cosmological constant, dynamic DE models, high energy fields, time-dependent cosmological constant, phase transitions, time-dependent gravitational coupling, quantum gravity effects, interacting dark matter-DE models, and various modifications of gravity theory beyond General Relativity.
In this study, the approach to evaluating these models centers on physics-agnostic, phenomenological representations
of time-varying equations of state. 
A straightforward, linear, two-parameter EOS offers enough detail to encompass a 
broad spectrum of models, as demonstrated, for example in \citet{linder2004}. 

The use of Type Ia SN data offers an opportunity to test dynamic DE models and determine which ones can better explain the observations, potentially alleviating the Hubble tension. 
Moreover, distinguishing among different EOS models provides valuable insight into 
alternate theories of gravitation, high energy physics, and higher dimensions, as noted by \citet{Linder_2003}.
However, efforts to constrain DE models using SN data are hampered by the accuracy and redshift coverage limitations
of the available data.
In the testing of alternative models using cosmological data,
the performance of statistical inference methods becomes crucial for evaluating
dynamic DE models. 
In cases where 
the model parameters exhibit strong degeneracies, traditional
$\chi^2$ minimization procedures often result in  
confidence regions that are excessively large, diminishing the ability to differenctiate between DE models.
To address this challenge 
we developed a more accurate statistical approach for model testing, 
which demonstrates reduced bias errors and enhanced statistical power.

The objectives of the current study are as follows:
(i) Investigate the extent to which dark energy models with time-varying equation of state can resolve the Hubble tension, 
with particular emphasis on a model where the EOS varies linearly with the scale factor $a$.
(ii) Test this model using Type Ia SNe data sourced from the Pantheon+ compilation. 
(iii) Determine confidence regions in the parameter space to explore alternatives to the $\Lambda$ cosmology and assess the statistical power for discriminating between models given the available SN data.
(iv) Given the challenges posed by parameter degeneracy and noisy data, develop a more robust statistic that enhances  discriminatory power beyond what is achievable with traditional $\chi^2$ methods.

In section~\ref{sec:basics} the fundamental theoretical background is presented.
Section~\ref{sec:analysis} delves into the testing of dynamic DE models using SN data.
Section~\ref{sec:cmbtest} introduces a consistency check with CMB results.
The discussion of the outcomes and conclusions is presented in section~\ref{sec:conclusions}.

\section{Basic Equations and Models}
\label{sec:basics}

The expansion rate of the Universe is defined by the Hubble parameter $H \equiv \dot{a}/a$. From the Friedmann equation,
it follows: 
\begin{equation}
\label{eq:exphistory}
   H(z) = H_0 \sqrt{\Omega_{r} (1 + z)^4 + \Omega_{M} (1 + z)^3 +\Omega_{k} (1 + z)^2 + \Omega_{DE} f(z)}
\end{equation}
The $\Omega_{x}$ terms are the standard density parameters for radiation, $r$, matter, $M$, curvature, $k$, and dark energy, $DE$.
The $f(z)$ function represents the evolution of the dark energy density.
For $H(z)$ and distance computations we followed the implementation of \citet{Wright_2006},
which sets $\Omega_{r}h^2 = 2.477 \times 10^{-5} (T_0/2.72528)^4$, with $h = H_0/(100$ km s$^{-1}$ Mpc$^{-1})$ and $T_0$ denoting the CMB temperature.

For a spatially flat-geometry ($\Omega_k = 0$) and allowing for dynamic dark energy,  
the comoving radial distance, $D$, to a source at a given redshift, $z$, is given by:
\begin{equation}
\label{eq:comdistance}
   D(z) = \frac{c A(z)}{H_0}
\end{equation}
with $A(z)$ defined as,
\begin{equation}
\label{eq:integral}
   A(z) = \int_{0}^{z}
   \frac{1}
    {\sqrt{\Omega_{r} (1 + z')^4 + \Omega_{M} (1 + z')^3 + \Omega_{DE} f(z')}}  dz'
\end{equation}
The equation for distance is expressed in terms of the function $A(z)$ to facilitate 
the discussion of how dynamic EOS models could address the Hubble tension. 
Specifically, representing distance as a ratio, $A/H_0$, illustrates the interaction 
between the parameters in the fit. For instance, focusing on a particular supernova in 
the sample at redshift $z$ and with magnitude $m$, the fit algorithm adjusts the model parameters
(which affect the numerator $A$ = $A(z;\Omega_M,w_0,w_a)$) and the parameter $H_0$ in the denominator
to bring the ratio $A/H_0$ as close as possible to the observed data $m$, where 
$m$ (SN magnitude) represents the logarithm of distance (Equation~\ref{eq:mag}). 
In the context of the 
Hubble tension, this interaction between the numerator and denominator in the distance equation implies that the high $H_0$ values could 
result from models overestimating the term $A(z)$ in the distance equation. 
If dynamic EOS models can decrease the values of $A(z)$ while
maintaining consistency with the low-redshift data, these models
could be promising candidates for alleviating the Hubble tension.

In flat geometry, the luminosity distance $d_L$ and the comoving radial distance, $D$, are related as follows: 
\begin{equation}
\label{eq:lumdistance}
   d_L = (1 + z) D 
\end{equation}

The $f(z)$ term in Equation~\ref{eq:integral} serves as a scaling model dependent factor for dark energy.
For the standard \lcdm\ model, $f = 1$.
The DE component can introduce a departure from the standard \lcdm\  model though $f(z)$,
potentially leading to
late-time acceleration increases that might address the $H_0$ tension. 
However, because the distance calculation involves an integral over $z$, the specific details of any DE model 
get averaged out, making it challenging to distinguish various DE models using distance data alone.

The relationship between the apparent magnitude (flux), $m$, the absolute magnitude (luminosity), $M_B$, and the luminosity distance, $d_L$ is given by: 
\begin{equation}
\label{eq:mag}
  m = 5 \log(d_L) + M_B + 25
\end{equation}
where $d_L$ is distance in Mpc units. 

The connection between data and models is established through the distance relations (Equations~\ref{eq:comdistance} and
\ref{eq:lumdistance}). 
In Equation~\ref{eq:comdistance} the function $f(z)$ encapsulates the dynamic component due to the EOS varying over time. 
To compute distance, it is necessary to specify an explicit model for the EOS function, $w(a)$. 
Various parametrizations have been proposed, including
$w(z)$ following an exponential function of the form $\exp{z/(1 + z)}$ by
\citet{pan2020}, and 
other parametrizations involving 
basic transcendental functions, such as $\log$, $\exp$, $\sin$, and 
$\arcsin$ \citep{pan2019}. The analysis conducted in this study adopts  
the Chevallier-Polarski-Linder (CPL) parametrization \citep{CPL,Linder_2003}, given by
\begin{equation}
\label{eq:cpl}
   w(a) = w_0 + w_a (1 - a) 
\end{equation}
where $a$ represents the cosmological scale factor, related to $z$ as  $a = 1/(1 + z)$.
In this parametrization,
the parameter $w_0$ denotes the present value of the EOS, 
and  $w_a$ determines the shape (negative slope) of the EOS variation with $a$.
The standard \lcdm\ cosmology corresponds to $w_0 = -1$, and $w_a = 0$.
The function $f(z)$ in Equation~\ref{eq:integral} takes the following form for the CPL parametrization \citep{Linder_2003},
\begin{equation}
\label{eq:fz}
   f(z) = (1+z)^{3(1+w_0+w_a)}  \exp\left(- \frac{3 w_a z}{1+z}\right)
\end{equation}

While the CPL parametrization is explicitly rooted in the equation of state of
a physical fluid (dark energy), models that directly influence the geometry can utilize
the simple, linear, phenomenological parametrization proposed by CPL,
provided that the mapping remains a reasonable approximation. 
In the literature, the 2D parameter space \wa\ has been employed to classify models based
on their location relative to the cosmological constant $(-1,0)$ model.
DE models that involve new high energy fields, often referred to as quintessence, 
fall into the $w < -1$ region, 
known as the phantom region. Conversely, introducing extra dimensions tend to 
occupy the $w > -1$ region. 

From the perspective of model testing, one of the advantages of CPL parametrization is that,
as long as theories can be mapped to a point on the 2D parameter plane \wa, 
models can be readily tested against data constraints, as demonstrated in section~\ref{sec:comparison}. 
~\citet{linder2004} shows that the CPL parametrization effectively encodes several models of interest
including,  
(i) Braneworld: Extra dimensions modify the Friedmann equation, reducing gravity in our
4D-brane \citep{Deffayet_2002};
(ii) Vacuum metamorphosis (Phantom) model: Expansion causes the quantum vacuum to undergo a phase 
transition at a redshift $z_j$, deviating from the matter dominated behavior \citep{Parker_2000};
And two models introduced by \citet{linder2004} as examples of acceleration modeled directly in the spacetime geometry:
(iii) SUGRA model: Characterized by a time-varying equation of state placing it on the $w > -1$ 
side of the 2D \wa\ plane; 
(iv) Ricci Geometric Dark Energy: Produced by a time-varying
normalized Ricci scalar curvature. These models serve as test cases for the statistical analysis 
presented in section~\ref{sec:analysis}.

\section{Probing the Expansion History with SN Data}
\label{sec:analysis}

The expansion history, $H(z)$, as described in Equation~\ref{eq:exphistory}, 
depends on the behavior of the dark energy EOS over time. 
If DE is the physical mechanism responsible for the acceleration of expansion, 
specific patterns in the expansion history - such as late time excess acceleration or a shift in the onset of acceleration -
could influence the $H_0$ tension. 
For example, if the onset of acceleration occurs at a late time, and the slope $dH(z)/dz$ deviates significantly
(in comparison to the nominal $\Lambda$ behavior) at late times, 
then the Hubble parameter derived from CMB observations will not align 
with that derived from luminosity-redshift data at low redshifts. 
The onset of acceleration and the magnitude of the late-time acceleration slope are determined by the EOS.

In R22, the \emph{Supernovae and $H_0$ for the Equation of State of dark energy}
(SH0ES) team presented the results of an analysis involving a subset of SNe from the Pantheon+ compilation.
In this analysis a $\chi^2$ minimization to a model that incorporates calibration parameters (SN Ia fiducial luminosity, Cepheid absolute magnitude, and Cepheid metallicity and luminosity parameters) results in a value of $H_0 = 73 \pm 1$. This value stands out as being $5\sigma$ away from the CMB derived value of 
$H_0 = 67.4 \pm 0.5$, giving rise to the Hubble tension. 
The SNe Ia selection used in R22 includes sources with redshifts up to $z < 0.15$,
whereas the Pantheon+ compilation encompasses SNe with redshifts up to $z < 2.3$.
As a result, the $H_0$ result derived by R22 can be characterized as local.

In a study by \citet{brout2022} the analysis is extended to larger distances ($z < 2.3$) encompassing
fits to various cosmological models,
including flat-$\Lambda$CDM, open-CDM, $w_0$-CDM and $w_{0}w_{a}$-CDM, where $w_{0}$, $w_{a}$ are EOS parameters
for an EOS linear in $z$, not in $a$ as in CPL.
The results of these fits yield
$H_0$ values with a variation among them within $<0.3\sigma$, 
indicating that SNe data ($z < 2.3$) alone lack sufficient statistical power to discriminate among alternative DE models. 
The topic of statistical significance of fits to SN data is further explored in the subsequent sections. 

\subsection{Fits to SN data}
\label{sec:fits}
The Pantheon+ sample, as described in \citet{pantheon},
consists of magnitude, redshift and the covariance matrix data for a total of 1701 SN Ia light curves 
(corresponding to 1550 distinct SNe). 
One of the key advantages of the Pantheon+ compilation is that it contains recalibrated SNe data from multiple photometric systems.
For this study, a subset of the Pantheon+ SN Ia compilation is utilized.
The analysis focuses specifically on the \emph{Hubble flow} (HF) region in redshift space, aiming to 
capture the expansion beyond perturbations caused by proper motions and the presence of void structures
in the local neighborhood. 
To generate the HF sample, selected SNe have $z > 0.15$, 
resulting in a total of 875 SNe with redshift extending up to $z < 2.26$.

The inference of model parameters, 
$H_0$, $\Omega_M$, $w_0$ and $w_a$, is performed using both, a least squares (LSQ) $\chi^2$ minimization approach, 
and using the new statistic \ta\ as described in Section~\ref{sec:ttest}.
The $\chi^2$ statistic is given by:
\begin{equation}
    \chi^2 = \mathbf{R}^T \mathbf{C}^{-1} \mathbf{R}
\end{equation}
where $\mathbf{R}$ is the residual vector, and $\mathbf{C}$ is the covariance matrix. 
The residuals are computed as follows: 
\begin{equation}
\label{eq:residuals}
   R_i = m_i - \left[ 5 \log(d_L(z_i;H_0,\Omega_M,w_0,w_a)) + M_B + 25 \right]
\end{equation}
Here, $m_i$ and $z_i$ represent the input data, where
$m_i$ is the peak apparent magnitude for the \emph{i-th} SN in the sample (corrected and standardized, Pantheon+ data item m\_b\_corr), $M_B$ is the fiducial SN luminosity, set as $M = -19.253$ based on R22, and $d_L$ is the luminosity distance as given in 
equations~\ref{eq:lumdistance}, \ref{eq:cpl} and \ref{eq:fz}. Notably, $d_L$
depends on the model through these equations. 
To be consistent with Planck-2018 CMB results, 
the model assumes flat geometry ($\Omega_k = 0, \Omega_{\Lambda} = 1 - \Omega_{M}$). 
The Covariance matrix, $\mathbf{C}$, is included in the Pantheon+ data release.
However, for use in the $\chi^2$ 
optimization code, the matrix is trimmed to account for the $z_{min}$ cut applied to the data. 

The parameters $M_B$ and $H_0$ are degenerate when analyzing SNe alone. This can be seen by 
rewriting equation~(\ref{eq:residuals}) using equations~(\ref{eq:comdistance}) and~(\ref{eq:lumdistance}) to 
expand $d_L$:
\begin{equation}
\label{eq:degeneracy}
   R_i = \left[m_i - 5\log(cA(z)(1 + z))\right] + \left[M_B + 25 -5\log(H_0)\right]
\end{equation}
where the sum of three terms in the second square bracket is a constant, hence the degeneracy. 
To break the $H_0/M_B$ degeneracy we fix $M_B$ equal to the SH0ES calibrated value, $M_B = -19.253$. This choice is driven by the high reliability and accuracy of the SH0ES $M_B$ value, calibrated based on a three-rung distance ladder utilizing precise \emph{Gaia} EDR3 parallaxes and distances to Cepheids based on the luminosity-period relation~\citep{Riess2022}.

\begin{figure}[h]  
   \centering
   \includegraphics[width=0.9\textwidth]{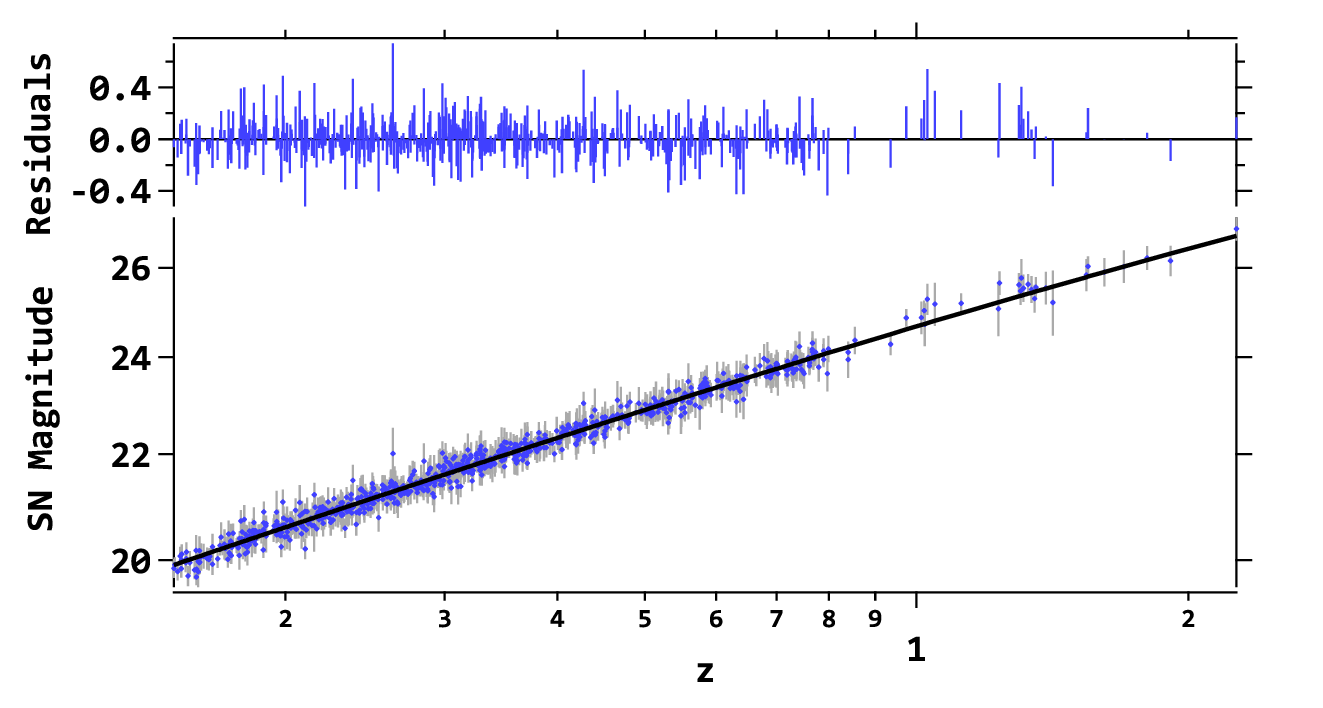}
   \caption{SNe magnitude data (blue dots), best fit (black) and residuals.}
   \label{fig:fit}
\end{figure}

The fitting process utilized a numerical optimization code incorporating a \emph{line-search} step method and 
the \emph{finite-differences} method for calculating the Hessian. The parameters subjected to fitting 
were $H_0$, $\Omega_M$, $w_{0}$ and $w_{a}$. 
The radiation density term, $\Omega_r$, was not considered as a fit parameter, instead,
it was fixed by the temperature of the CMB.
The uncertainties associated with the fit parameters were determined through a Monte Carlo procedure,
outlined below,
and solely reflect statistical errors.
The results of the fit are presented in Table~\ref{tab:fit} and illustrated in Figure~\ref{fig:fit}.

\begin{table}
   \caption{Fit of the CPL dynamic dark energy model to Pantheon+ SNe data.
   Parameter errors (in the `Fit' column) represent the $84th$ and $16th$ percentiles of the 
   Monte Carlo-generated marginalized distributions of the differences $p_{true} - p_i$ 
   (where $p_i$ represents the parameter value on the $i-th$ realization and $p_{true}$ is the true value of the parameter).
   The units for $H_0$ are \ksmpc.
   The `StdDev' column shows the 
   standard deviation of the marginalized distributions.
   Bias error is the difference between the mean value (from the Monte Carlo-generated distributions) 
   and the true parameter value. 
   The results for $w_0$ and $w_a$
   displayed below $T_{\alpha}$ are obtained using the $T_{\alpha}$ statistic (see section~\ref{sec:ttest})
   PTE denotes the probability to exceed, calculated from a $\chi^2$ distribution with $N_{dof} = 871$ degrees of freedom.
   }
   \label{tab:fit}
   \begin{tabular}{@{}lccc@{}}
   \toprule
   Parameter  &   Fit  & StdDev & Bias Error \\
   \midrule
        $H_0$        &  $74.4^{+1.4}_{-4}$        &  $2.9$   & -1.3 \\
        $\Omega_{M}$ &  $0.275^{+0.193}_{-0.071}$ &  $0.129$ & 0.06 \\
        $w_0$        &  $-1.1^{+0.8}_{-0.3}$      &  $0.77$  & 0.3 \\
        $w_a$        &  $1.2^{+0.6}_{-7.6}$       &  $5.6$   & -3  \\
   \midrule
        $T_{\alpha}$ minimization &               &          &  \\
        $w_0$        &  $-1.2 \pm 0.6$            &  $0.7$   & -0.03 \\
        $w_a$        &  $2 \pm 2.7$               &  $2.9$   & 0.07  \\
   \midrule
        $\chi^2_{min}$ & $775$                    &          & \\
        $N_{dof}$    & $871$                      &          & \\
        $PTE$        & $0.9912$                   &          & \\
   \botrule
   \end{tabular}
\end{table}

\subsection{Monte Carlo}
\label{sec:montecarlo}
To generate the probability distributions of the fitted parameters, a Monte Carlo was implemented following the Ordinary Monte Carlo (OMC) procedure. Although in current astrostatistics practice Markov Chain Monte Carlo (MCMC) is more commonly used, in the case of dynamic dark energy models, OMC offers key advantages. The characteristics of the DE scaling function, $f(z)$ in Equation~(\ref{eq:fz}), are such that the joint posterior distribution of parameters \wa\ is highly elongated and exhibits two regions of low $\chi^2$ residuals (one along positive $w_a$, the other along negative $w_a$). These characteristics present challenges for MCMC algorithms, specifically in defining appropriate priors and choosing adequate samplers so that the algorithm does not get stuck inside a limited region of parameter space. At the expense of a modest loss of efficiency, a brute-force OMC completely spans the parameter space, offering a more robust solution. In addition, OMC methods support the task of assessing the accuracy of parameter fitting (section~\ref{sec:ttest}) and exploring data accuracy requirements for future experiments (section~\ref{sec:comparison_discuss}).

The OMC procedure was implemented as follows: 
(i) the values of the fitted parameters (Table~\ref{tab:fit}) were employed as proxies 
for the true values of the model parameters; 
(ii) 5000 random realizations of synthetic data sets were generated
based on these true parameter values; 
(iii) for each simulated data set, the $\chi^2$ and $T_{\alpha}$ optimization code 
were executed to obtain the best fit parameters for each realization. 
The distributions of the best fit parameters
from each realization provide the desired information. 
Each synthetic data set consists of a vector of
875 values (matching the sample size of the main fit) of SN magnitudes generated using the model equations (Equations~\ref{eq:comdistance}, \ref{eq:lumdistance}, 
\ref{eq:mag}, \ref{eq:cpl},
and~\ref{eq:fz}) evaluated at the corresponding redshifts ($z_i$) of each SN, 
along with Gaussian random noise with $\sigma = 0.2 mag$ (representing the average magnitude error of the Pantheon+ sample).

\begin{figure}[h]
   \includegraphics[width=\columnwidth]{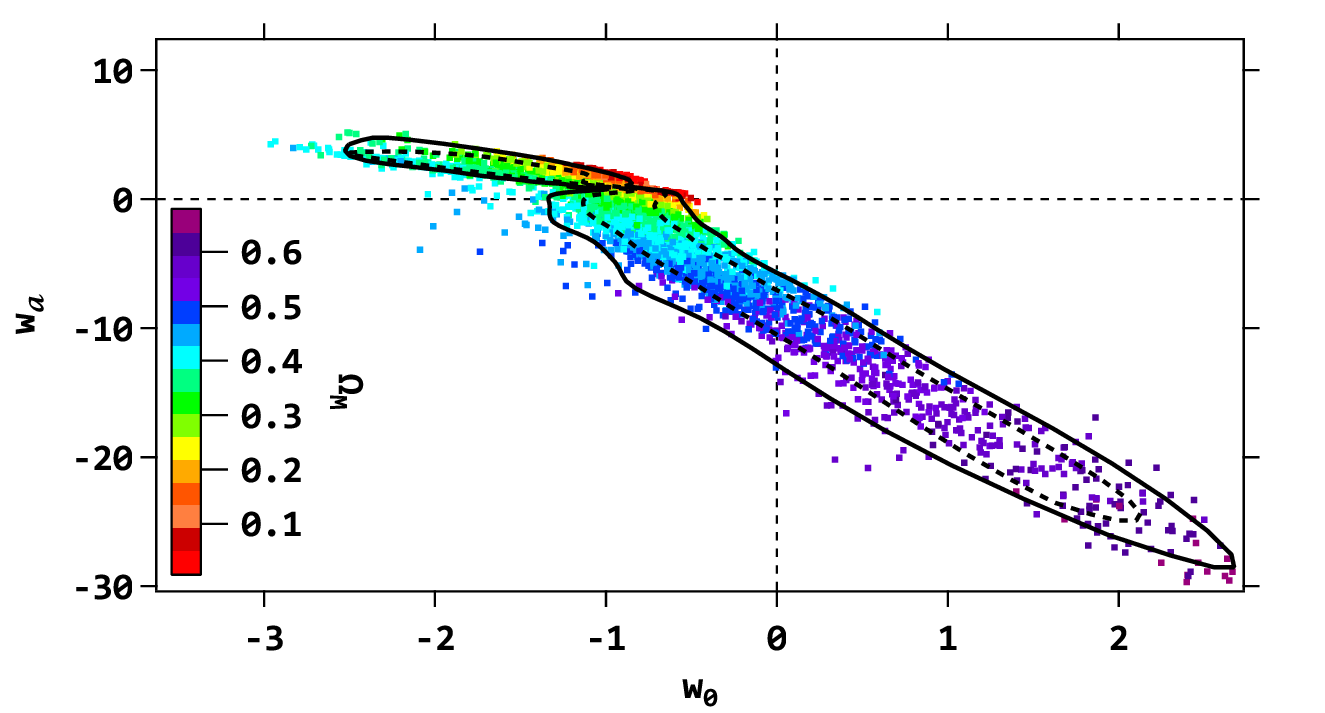}
   \caption{Joint distribution of Mote Carlo generated EOS parameters $w_0$ and $w_a$ using $\chi^2$ minimization.
   The 95\% CL (solid black line) and 68\% CL (broken line) contours show large parameter degeneracy.
   The color scale represents the values of parameter $\Omega_M$.
   } 
   \label{fig:mc_contours_omega}
\end{figure}

\begin{figure}[h]
   \includegraphics[width=\columnwidth]{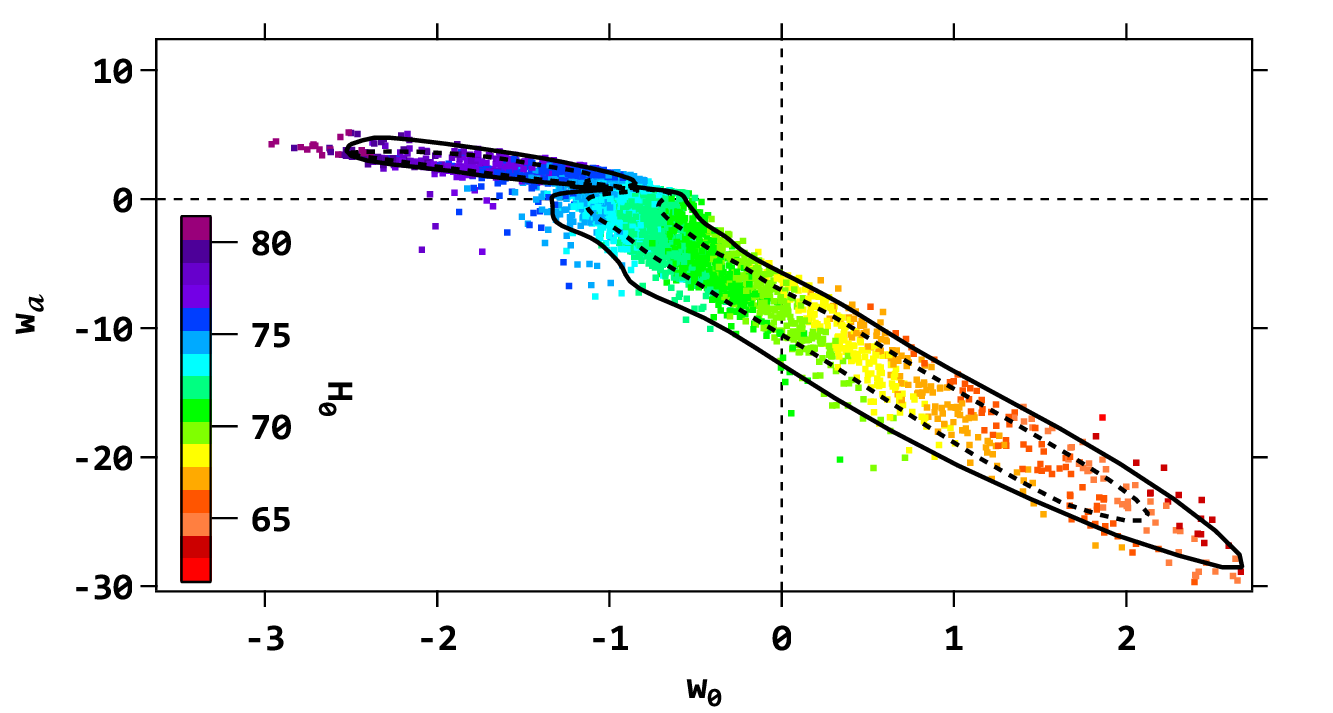}
   \caption{Joint distribution of Mote Carlo generated EOS parameters $w_0$ and $w_a$, same as Figure~\ref{fig:mc_contours_omega} 
   but with the color scale representing the values of parameter $H_0$.
   }
   \label{fig:mc_contours_h0}
\end{figure}

\begin{figure}[h]
   \includegraphics[width=\columnwidth]{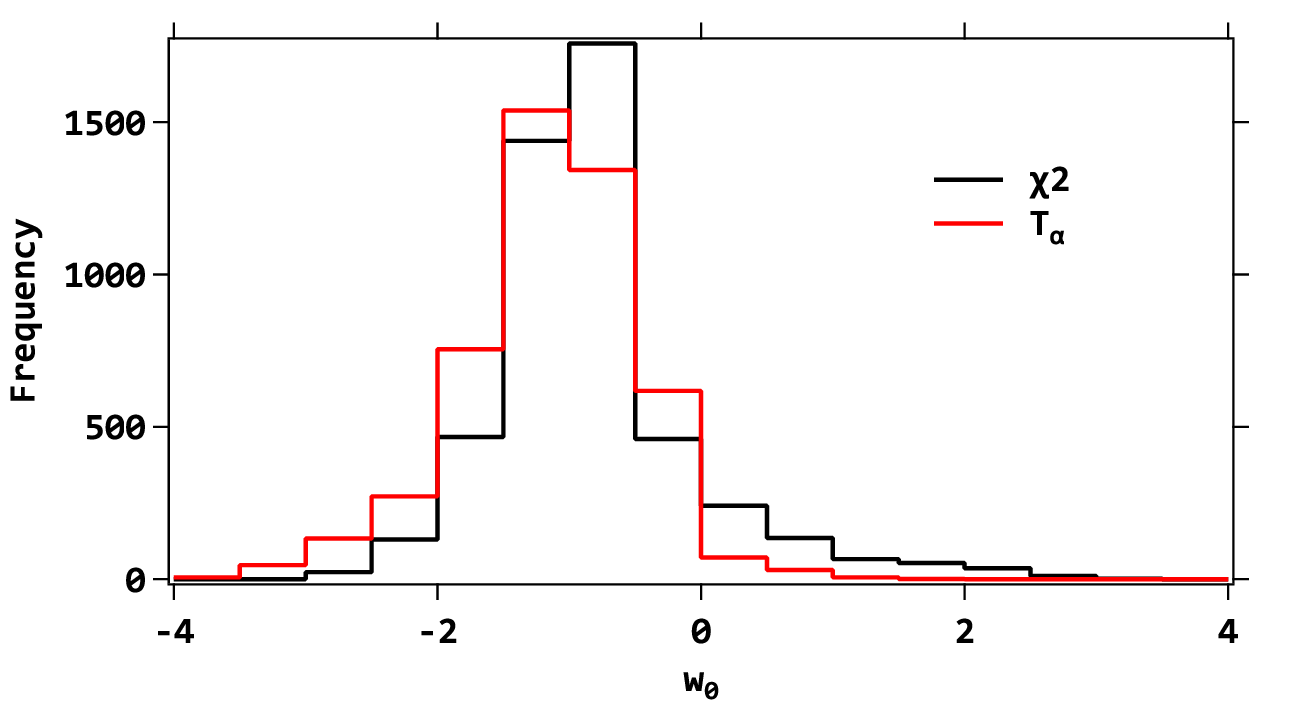}
   \caption{Marginalized distribution of Mote Carlo generated EOS parameter $w_0$ 
   using $\chi^2$ (black) and $T_{\alpha}$ (red) minimization.}
   \label{fig:histo_w0}
\end{figure}

\begin{figure}[h]
   \includegraphics[width=\columnwidth]{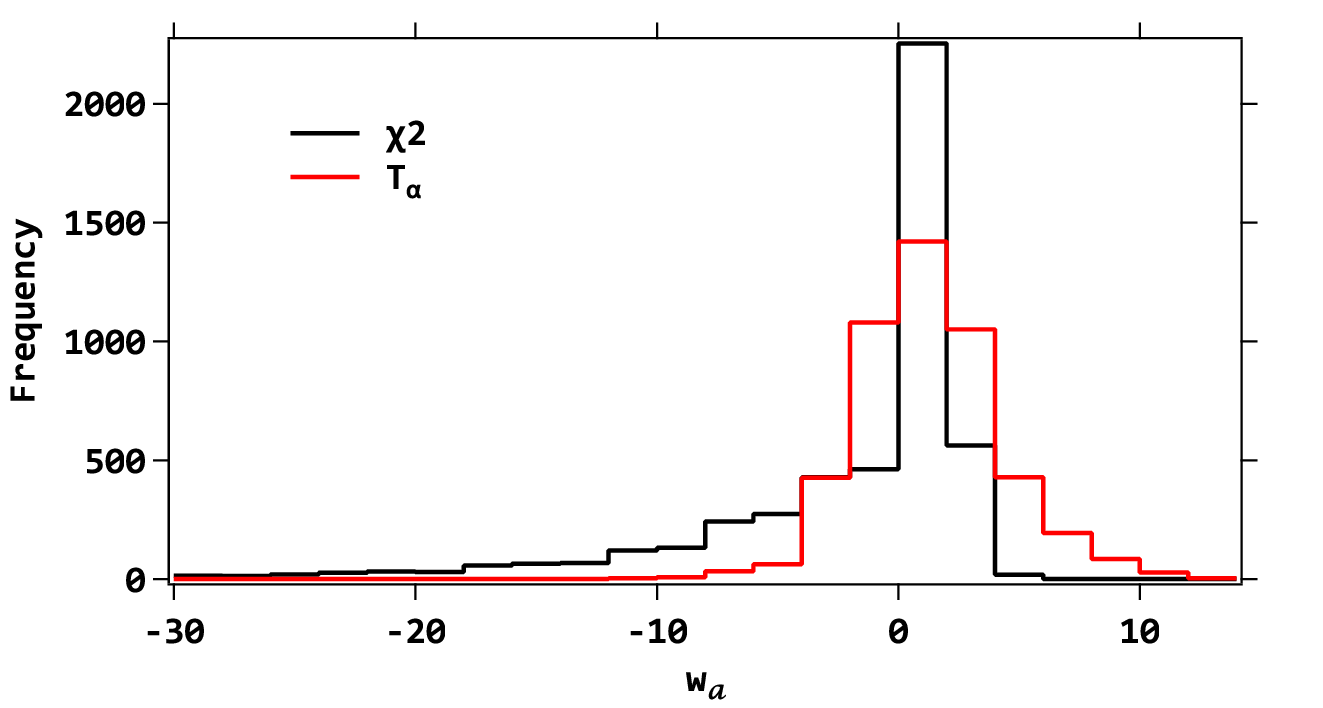}
   \caption{Marginalized distribution of Mote Carlo generated EOS parameter $w_a$ 
   using $\chi^2$ (black) and $T_{\alpha}$ (red) minimization.}
   \label{fig:histo_wa}
\end{figure}

The joint distributions for the CPL parameters are visualized in figures~\ref{fig:mc_contours_omega} and \ref{fig:mc_contours_h0}.
Marginalized distributions for parameters $w_0$ and $w_a$, presented separately, 
can be observed in figures~\ref{fig:histo_w0}
and \ref{fig:histo_wa}. 

\subsection{Discussion of Fit Results}
\label{sec:fit_discussion}
Upon inspecting
the fitted curve against the data (Figure~\ref{fig:fit}), it is evident that the 
fit is reasonably good. However, in this case 
the $\chi^2$ statistic cannot be used as a measure of goodness-of-fit. 
Considering that the mean of the $\chi^2$ statistic is equal to $N_{dof} (871)$,
a $\chi^2_{min}$ value of $775$, less than $N_{dof}$, 
suggests that the measurement errors might be overestimated, potentially due to covariance terms.
Whereas the 
elements of the covariance matrix along the diagonal are always positive, 
cross-variance terms could contribute to the $\chi^2$ with a negative sign.
To assess the quality of the fit (in the absence of a reliable $\chi^2$), it's worth noting that 
the RMS of the residuals ($0.147 mag$) aligns with, or is even lower than, the average magnitude error ($0.22 mag$).
This consistency implies a good agreement between the fitted model and the data. 

\subsubsection{Parameter Errors}
Due to the inclusion of systematic errors in the Pantheon+ covariance matrix,
the fit results for $H_0$ (Table~\ref{tab:fit}) incorporate systematic errors as well.
When using the $\chi^2$ statistic in the fits, the parameter distributions become skewed, 
causing the mean to be displaced relative to the true values. 
This skewness results in (i) bias errors, and  (ii)
non-symmetric errors: $\sigma _{+} = 1.4$ and $\sigma _{-} = 4$, for $H _{0}$.
These errors are higher than the $H_0$ error reported by R22 ($\pm 1.04$).
The increase in errors can be attributed to the addition of two degrees of freedom to the fit 
($w_0$ and $w_a$) and the expansion of the redshift range ($z < 2.26$).
 
When using the $\chi^2$ statistic,
the distributions of the CPL parameters \wa\ exhibit similar skewness
(as evident in figures~\ref{fig:histo_w0} and~\ref{fig:histo_wa}).
However, this skewness is not observed when employing the $T_{\alpha}$ statistic.
It is noted that the 68\% and 95\% confidence contours in the \wa\ parameter space
(figures~\ref{fig:mc_contours_omega} and~\ref{fig:mc_contours_h0})
demonstrate significant parameter degeneracy
spanning a broad region. 
Specifically,
$w_0$ ranges from $-2.8$ to $+2.6$ and $w_a$ ranges from $-27$ to $+4$.
This wide parameter space diminishes the statistical power 
to either reject or confirm alternative DE models.

\subsubsection{Parameter Degeneracy}
The confidence regions in the \wa\ parameter space exhibit a high degree of parameter degeneracy,
forming two elongated lobes.
One lobe  primarily extends along positive $w_a$, following the line $w_a = -1.68 w_0 - 0.75$,
while the other longer lobe along negative $w_a$ has a steeper
negative slope following the line $w_a = -7.4 w_0 - 7.9$. 
The influence of the \wa\ parameters in the $A(z)$ term (Equation~\ref{eq:integral}) is such that 
combinations of $w_0$ and $w_a$ along the degeneracy lines tend to yield similar 
values for $A$. 

\begin{figure}[h]
   \includegraphics[width=\columnwidth]{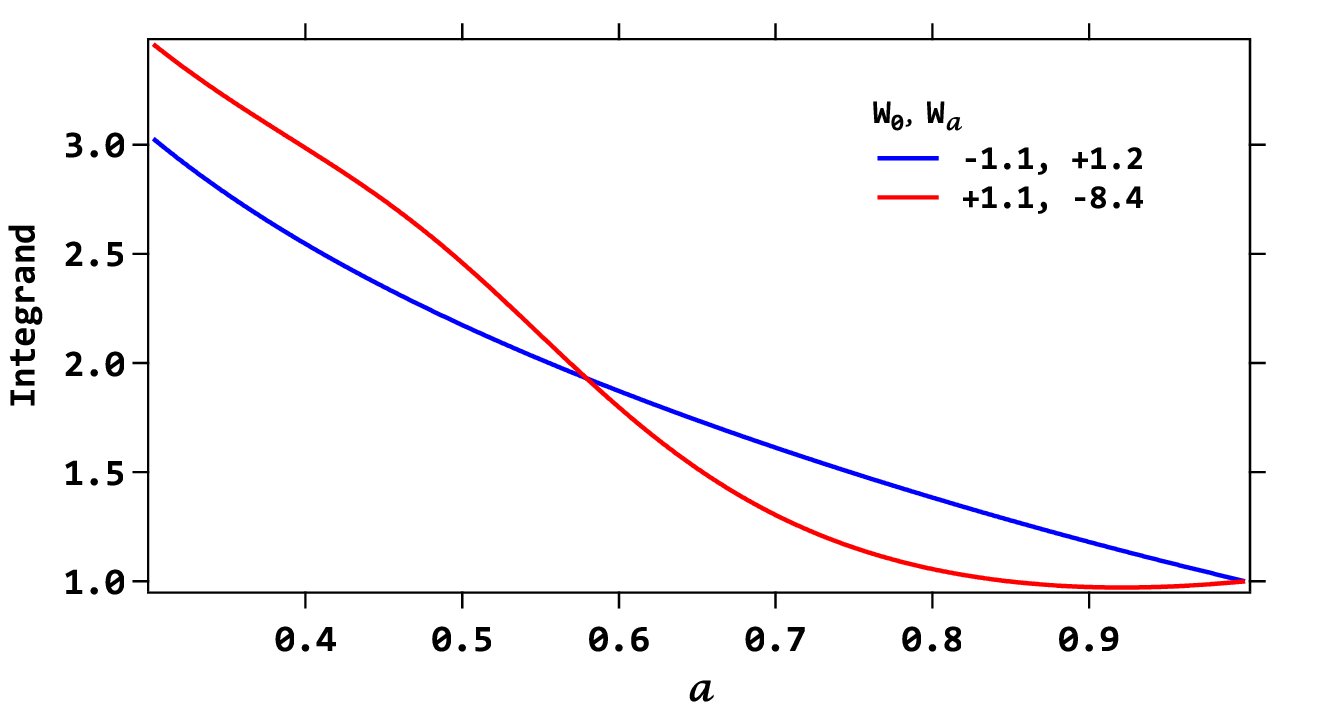}
   \caption{Function (after a change of variable from $z$ to $a$) inside the integral of the term $A(z)$
   Equation~\ref{eq:integral}
   for two settings of the \wa\ pair laying along
   a degeneracy axis. This example corresponds to a source at $z = 2.3$.}
   \label{fig:integrand}
\end{figure}

The effect of parameter degeneracy is illustrated in Figure~\ref{fig:integrand}, which shows curves
of the function inside the integral in $A(z)$ for a source at $z = 2.3$, considering two sets
of \wa\ parameters along a degeneracy axis. 
It can be observed that the area under the curves is the same for both settings of the \wa\ parameters,
indicating that the distances (and SN magnitudes) computed for this particular source ($z = 2.3$) 
are identical.
For sources with redshifts $z < 2.3$ the area under the curve would differ depending on the parameter set, 
however,
the differences in magnitude are small compared to the data errors.

In addition to the degeneracies among the CPL parameters, there are degeneracies with 
the cosmological parameters $H_0$ and $\Omega_M$, as indicated by the color scale on the
Monte Carlo points in figures~\ref{fig:mc_contours_omega} and~\ref{fig:mc_contours_h0}. 
The $\Omega_M$ degeneracy pattern exhibits points with low $\Omega_M$ (red) toward the positive
$w_a$ side of the confidence region, and high values (blue) toward the negative
$w_a$ values. The $H_0$ degeneracy pattern shows higher 
values (blue) on the positive $w_a$ side and lower values (red) toward the negative $w_a$ side. 
This intricate pattern of multi-parameter degeneracies helps elucidate the observation that 
dynamic DE (at least in the CPL formulation) has a marginal impact on alleviating the 
Hubble tension. 

The CPL mechanism operates in the following manner:
(i) The \lcdm\ model tends to overestimate the distances (and consequently, magnitudes)
of SNe in the HF region ($0.15 < z < 2.26$);
(ii) A fit to the SN data using the \lcdm\ model would therefore overestimate $H_0$ (in order to 
maintain the $A(z)/H_0$ ratio consistent with the data, as explained in the  introduction);
(iii) A dynamic DE model with \wa\ parameters skewed toward positive $w_a$ values 
would decrease the estimated distances at these redshifts, leading to 
a lower $H_0$. This logic implies that \emph{if} the true universe
followed a cosmology with a dynamic dark energy equation of state parametrized according to CPL, \emph{then}
the measured $H_0$ would be lower than the high value obtained from fits to the \lcdm\ model. 

However, this is not what is observed. In fact, the numbers in Table~\ref{tab:fit} show that
$H_0$ does not decrease relative to the high $H_0$ reported in R22 ($73$ \ksmpc). 
This behavior is explained by noting that while the additional degrees of freedom \wa, 
allow, in principle, for a lower $H_0$, the degeneracy with the 
parameter $\Omega_M$ counteracts this effect by pushing the fit towards a lower $\Omega_M$. 
To illustrate, a fit to the same SN sample ($0.15 < z < 2.26$) while holding \wa\ to the 
\lcdm\ values ($-1,0$) yields $H_0$ = 73.4 and $\Omega_M = 0.33$, which is higher 
than the $\Omega_M$ obtained from the CPL fit ($0.275$). 

While dynamic DE models (with CPL parametrization) do not completely resolve the $H_0$ tension,
they are not entirely ruled out (as pointed out, the fit to SN data is reasonably good). 
They could potentially play a partial role 
in addressing the Hubble tension, specifically if they operate within 
a restricted interval in redshift space, as proposed in the sigmoid EOS
model in \citet{torres2023}. 
However, it is essential to understand why results from studies like R22, which focus on the local
measurement of $H_0$, yield higher values
even in a model-independent context.

In R22, the local measurement of $H_0$ is largely unaffected by the inclusion of dynamic DE models.
R22 uses a sub-sample of the Pantheon+ data in the low redshift region ($z < 0.15$), where
a simple low-redshift approximation for distance is appropriate. 
While R22's analysis involves a multi-parameter fit, 
with respect to $H_0$, it can be seen as a straightforward fit to a
magnitude-redshift relation of the form: $\log H_0 = 5 + \log cz + (M_B - m)/5$.

The focus of this work is on confronting dynamic DE models with SN data in the Hubble
flow region $z > 0.15$.
If the fit had yielded a lower $H_0$ value (compatible with CMB) 
then, despite the fit results, R22's model-independent results for the local $H_0$ would remain unaffected. 
This suggests that the high value of the local $H_0$ could be due to unaccounted for proper motions, or  
the model would need to incorporate a drastic transition at low-redshift that significantly reduces the 
estimated distances, 
which is not supported by dynamic EOS models because they rely on the Friedmann framework.
Dynamic DE models don't fundamentally affect the local $H_0$ determination because
$H_0$ is the anchor point (at $z = 0$) for the expansion history, $H(z)$,
independent of how DE models modify the $A(z)$ integral. 
The limited role that a dynamic DE model could play would 
need to be accompanied with yet another physics process that generates a drastic transition 
of the distance equation at low-redshifts (to lower the estimate of distances).

\subsection{The T-alpha  Statistic}
\label{sec:ttest}
\begin{figure}[h]
   \includegraphics[width=\columnwidth]{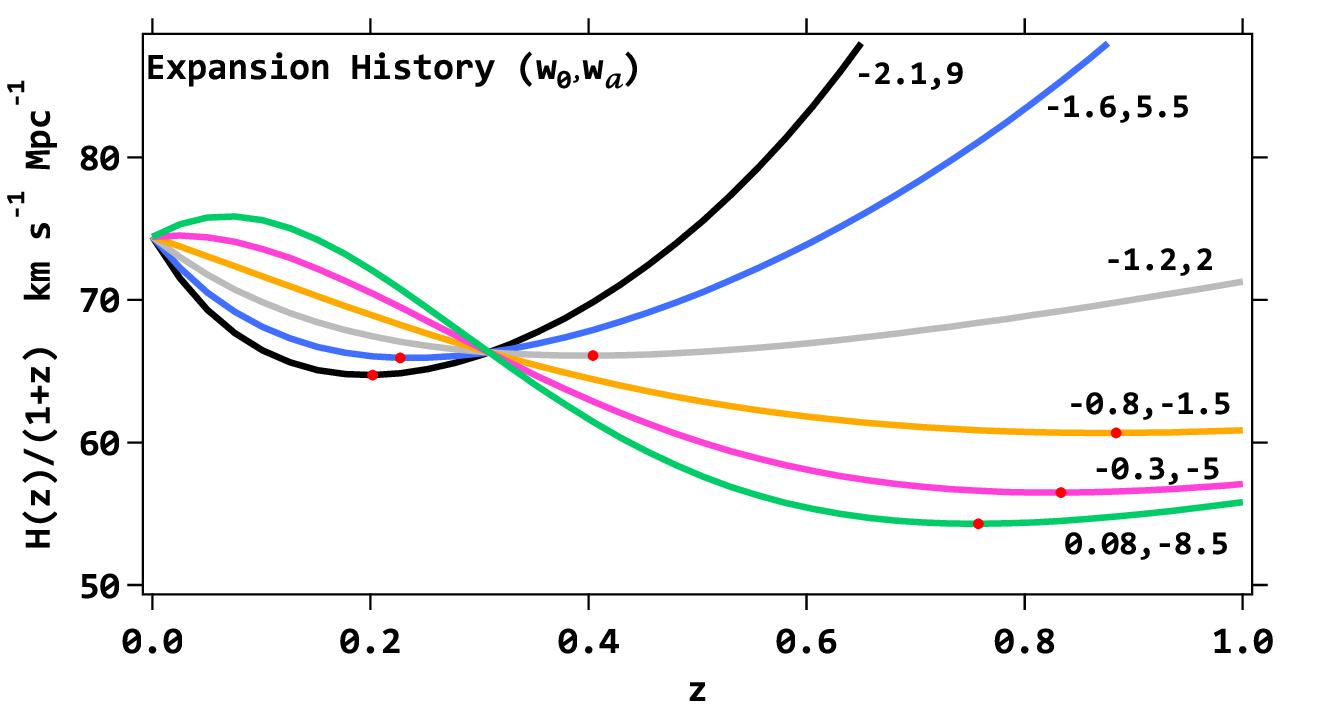}
   \caption{Comoving Hubble parameter for DE model variants along
   the main degeneracy axis. The onset of acceleration for each curve is indicated by red dots.
   Curves with $w_0 \geq -0.3$ (pink and green) show deceleration instead of acceleration at $z = 0$. }
   \label{fig:h_z}
\end{figure}

\begin{figure}[h]
   \includegraphics[width=\columnwidth]{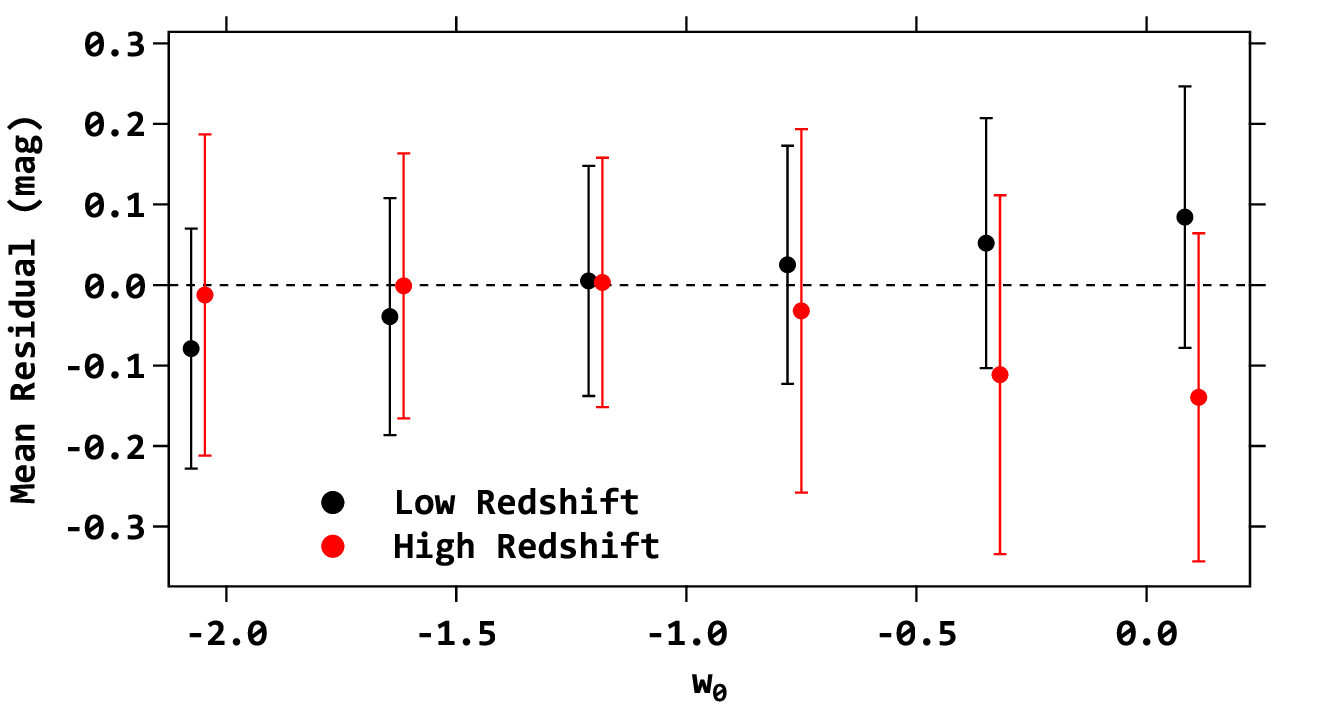}
   \caption{Mean of residuals for the set of $w_0$, $w_a$ models used in Figure~\ref{fig:h_z}
   showing the high/low redshift asymmetry. 
   For each model the means are separately computed for two bins: low redshift (black) and high redshift (red)
   separated at the redshift where the corresponding $H(z)/(1+z)$ curve reaches the minimum.
   The error bars represent the standard deviations of the residuals in the corresponding bin.}
   Note: for clarity, the red points have been shifted slightly to the right
   \label{fig:h_meanres}
\end{figure}

From the discussion of fit results it has been shown that using 
$\chi^2$ minimization to infer the parameters of the CPL parametrization results in 
large bias errors, strong parameter degeneracy, and large confidence regions, diminishing discriminatory power 
when comparing alternative models. To alleviate these issues we developed the \ta\ statistic, 
which involves comparing the model with data, taking into account the sign of the difference.
The advantage provided by the \ta\ 
stems from the fact that the $\chi^2$ statistic looses information (i.e. the sign of the residuals) when 
squaring the differences. However,
in comparing DE models using CPL parametrization, the sign of the residuals 
carry meaningful information, which the \ta\ statistic exploits. 
The residuals exhibit an asymmetric pattern depending on the \wa\ parameters:
when the model parameters are sub-optimal, the mean of residuals is positive biased
for low-redshift while negative biased for high-redshift, or vice versa.
The \ta\ statistic is sensitive to this pattern, in contrast, 
traditional \(\chi^2\) minimization, by squaring the differences, 
tends to treat positive and negative deviations symmetrically, potentially leading to loss of important information.
To illustrate how these patterns work, Figure~\ref{fig:h_z} displays
curves of the comoving Hubble parameter as a function of redshift for 
various combinations of \wa\ parameters along the main degeneracy axis. It is important to note 
that these curves have a ``U'' shape with a minimum point corresponding to 
the onset of acceleration. This ``U'' shape in the curves causes an asymmetry
in the sign of the residuals when the \wa\ parameters are sub-optimal. 
Figure~\ref{fig:h_meanres} shows the mean of residuals computed for redshifts to the left (black points) and the right (red points)
of the $H(z)/(1+z)$ minima of the \wa\ models used in Figure~\ref{fig:h_z}. 
The error bars represent the standard deviations of the residuals.
The high/low redshift asymmetry is apparent. 
Due to the significant variation in the $w_0$ and $w_a$ parameters along the degeneracy axis, 
the expansion history exhibits vastly different outcomes:
the onset of acceleration (minima, marked with red dots) ranges from $z = 0.2$ to $z = 0.88$, 
and the minimum of $H_0/(1+z)$ (which determines the acceleration slope at late times) ranges from $54$ to $66$ \ksmpc.
By balancing the residuals at both ends of the redshift range, the \ta\ statistic provides a more robust estimation of the parameters.

The \ta\ statistic is a \emph{T-test} that compares the mean of residuals 
computed in two regions of redshift space: low and high redshift, with the partition point determined 
by the parameter $\alpha$.
Minimizing this statistic
with respect to model parameters leads to a more accurate parameter estimation. 
The procedure consists of the following steps:
(i) The LSQ-$\chi^2$ minimization code (section~\ref{sec:fits}) is initially executed to obtain an initial set of parameter values;
at this point the LSQ-$\chi^2$ fitted values for $H_0$ and $\Omega_M$ are held constant while $w_0$ and $w_a$ vary during optimization (step iii below);
(ii) Compute the residuals vector (Equation~\ref{eq:residuals}) and sort it by redshift. 
Let $\alpha$ be an index referring to an arbitrary element in the sorted list of residuals,
and $z_{\alpha}$ the corresponding redshift in the list.
Define the function $T(\alpha)$ as the value of a \emph{T-test} comparing the means of residuals computed in two bins: 
one containing data points with $z \leq z_{\alpha}$,
and the other with points that satisfy $z > z_{\alpha}$. 
The T-test is defined as follows:
$T(\alpha) = (\bar{r}_H - \bar{r}_L)/S_D$,
where $\Bar{r}_x$ represents the mean of residuals computed in one of the bins, high-z bin (subindex H) and low-z bin (subindex L),
and $S_D$ is the \emph{standard error}: $S_D^2 = N (\nu_H \sigma^2_H + \nu_L \sigma^2_L)/N_{DOF}$
with $\sigma_x$ the standard deviation of residuals ($x$ denotes the bin, H or L, as before), $\nu_x = n_x - 1$, $n_x$ the number of points in the bin, $N_{DOF}$ degrees of freedom, $N_{DOF} = \nu_H + \nu_L$, and $N = (1/n_H + 1/n_L)$.
The T-test as defined above is applicable when the distributions in the high and low bins have the same (or nearly the same) standard deviation, which is the case in this analysis as shown by the error bars of similar length for the high and low redshift bins in Figure~\ref{fig:h_meanres}.

The \ta\ statistic is defined as the maximum absolute value of $T(\alpha)$ as $\alpha$ runs from $k$ to $N-k$:
\begin{equation}
\label{eq:ttest}
    T_{\alpha} = \max_{k \leq \alpha \leq (N-k) } \| T(\alpha) \|
\end{equation}
where $N$ is the number of points, and $k$ is a margin at the ends of the list to guard against bins with too few
data points. 
(iii) Run an optimization code to minimize the \ta\ statistic varying 
the values of parameters \wa\ such that they satisfy the constraint
$w_a = A w_0 + B$, 
where $A$ and $B$ are parameters that define a linear relation along the degeneracy axis.
To summarize, the $T_{\alpha}$ optimization algorithm is an incremental improvement of the fit results using $\chi^2$ minimization and operates only on 
the $w_0$, $w_a$ parameters while $H_0$ and $\Omega_M$ retain their $\chi^2$ fitted values.

\begin{figure}[h]
   \includegraphics[width=\columnwidth]{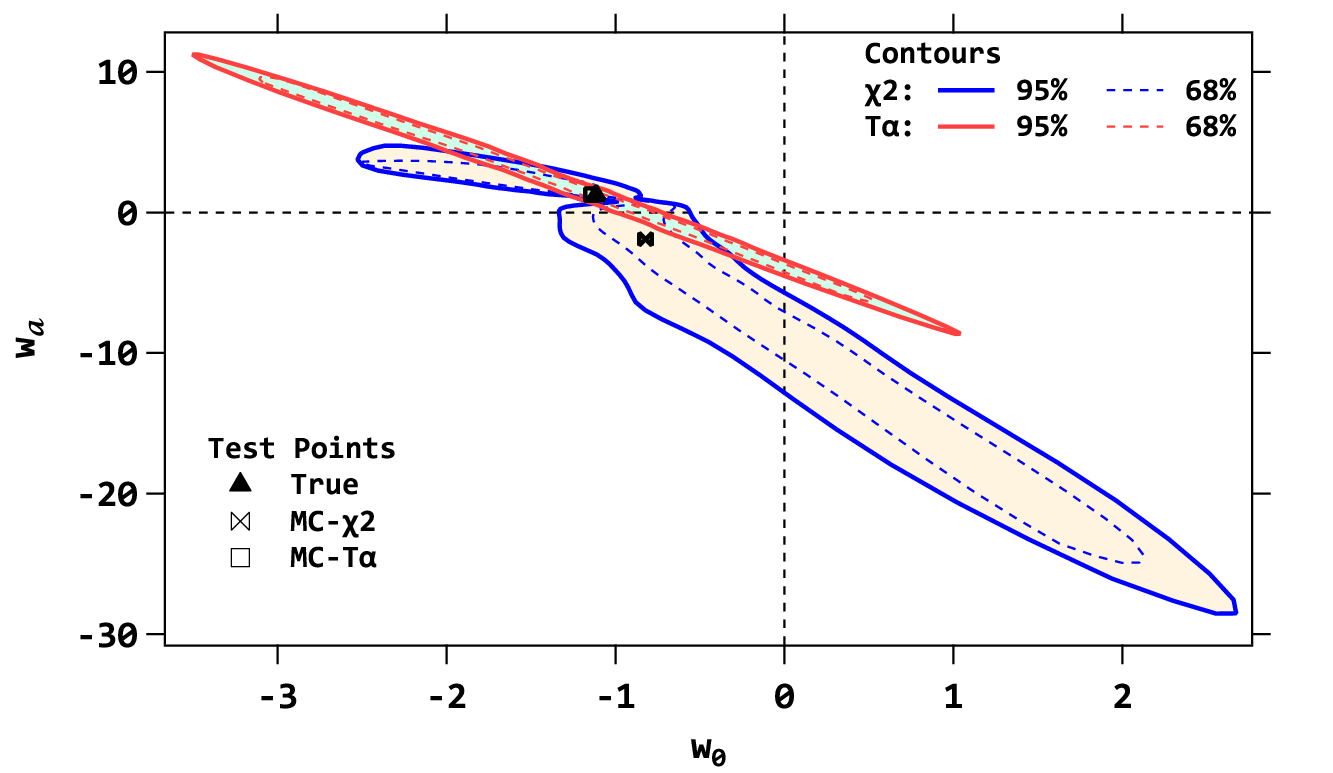}
   \caption{Monte Carlo generated 68\% and 95\% CL contours comparing the performance 
   of the 
   $\chi^2$ (blue curves) with the $T_{\alpha}$ (red curves) statistic.
   The points included in the plot are: true parameter values (solid triangle),
   average of Monte Carlo generated points using $\chi^2$ (two-triangle shape),
   and average using \ta\ (open square).  
   }
   \label{fig:contours_s0p2}
\end{figure}

The Monte Carlo procedure described in section~\ref{sec:montecarlo} included running the \ta\ minimization
fits. Figure~\ref{fig:contours_s0p2} shows the 68\% and 95\% CL contours 
resulting from \ta\ minimization, alongside the contours from $\chi^2$ minimization for comparison. 
The narrower \ta\ contours illustrate the advantages of the \ta\ statistic. Bias errors
can also be observed in the plot by noting the distance between the Monte Carlo averages 
(represented by the square and two-triangle shape) and the true parameters (depicted by the solid triangle).
The smaller bias error of \ta\ is evident by the position of the open square in the plot, which is barely 
visible as it is positioned right behind the solid-triangle (denoting the true values). 
The numeric values of bias errors are provided in Table~\ref{tab:fit}. 

\subsection{Model Comparison}
\label{sec:comparison}
In this section, we will address the following questions:
(i) What can we infer about the statistical significance of the CPL model fit?
(ii) Based on the available SNe data, can we exclude any of the alternative models mentioned in the introduction?
(iii) To what extent can the available SNe data differentiate between alternative DE models and the baseline \lcdm?
To conduct the necessary statistical tests, one approach is to compare the \wa\ parameters with one of two reference points.
This choice depends on the intended type of test: the reference point could be
either the baseline \lcdm\ (to establish the level of significance),
or the best-fit parameters (to assess compatibility with the data). The analysis presented below adheres to standard hypothesis testing methods (refer to, for instance, \citet{workman}).

When testing the significance of a signal amidst noise, the relevant examination aims to evaluate the probability that an alleged signal could merely be a chance occurrence due to noise alone. 
In this scenario, there is no actual signal, and the outcome represents
a false-positive error, also known as \emph{Type-I error}.  
The level of significance quantifies the risk of a Type-I error. 
A signal can be asserted as detected when the probability of Type-I errors is low (e.g., $< 1.3 \times 10^{-3}$ for $3\sigma$, $2.87 \times 10^{-7}$ for $5\sigma$, one-tailed). In our case, the `signal' refers to the presence of deviations in the dark energy EOS from the baseline \lcdm.

In the second case (comparison against best-fit results), 
the test can be used to rule out alternative models. 
If the best-fit model is reasonably good (e.g., the model is consistent with the data, 
according to some goodness-of-fit criteria), 
comparing an alternative model against the best-fit model could provide insights into the consistency of the alternative model with the data.

In the 2D \wa\ parameter space, model comparison is conducted by computing the normalized distance between the \wa\ point
to be tested and the reference point. 
Using the `signal' analogy, the strength of the signal is determined by this normalized distance,
which is adjusted based on the appropriate standard deviations, often referred to as `sigmas'. 
This sigmas are derived from the dispersion of the Monte Carlo points. 
However, since the Monte Carlo points result in a joint distribution aligned along a steep degeneracy axis,
the sigmas must be calculated along the principal axes.
To achieve this,
a principal component analysis (PCA) is performed to obtain the relevant sigmas for the test. 
This process involves rotating the axes so that the sample variances along the new axes represent the extremes (maxima and minima) 
and are uncorrelated.
Consequently, the distance between the model under test and the reference point is calculated using the following formula,
\begin{equation}
    \label{eq:nsigma}
    N_{\sigma} = \sqrt{\left( \frac{\Delta p_x}{\sigma_x} \right)^2 + \left( \frac{\Delta p_y}{\sigma_y} \right)^2}
\end{equation}
Here, $\Delta p_x$ and $\Delta p_y$ denote the differences between the model and the reference point 
expressed along the PCA rotated axes ($x,y$), and $\sigma_x$, $\sigma_y$
are the standard deviations of Monte Carlo points computed along the rotated axes.
Notably, $N_{\sigma}^2$ follows a $\chi$ squared distribution, $\chi^2_p$ with $N_{dof} = 2$. 
The associated probability to exceed (PTE) conveniently provides a measure 
of the level of significance. 

\begin{figure}[h]
   \includegraphics[width=\columnwidth]{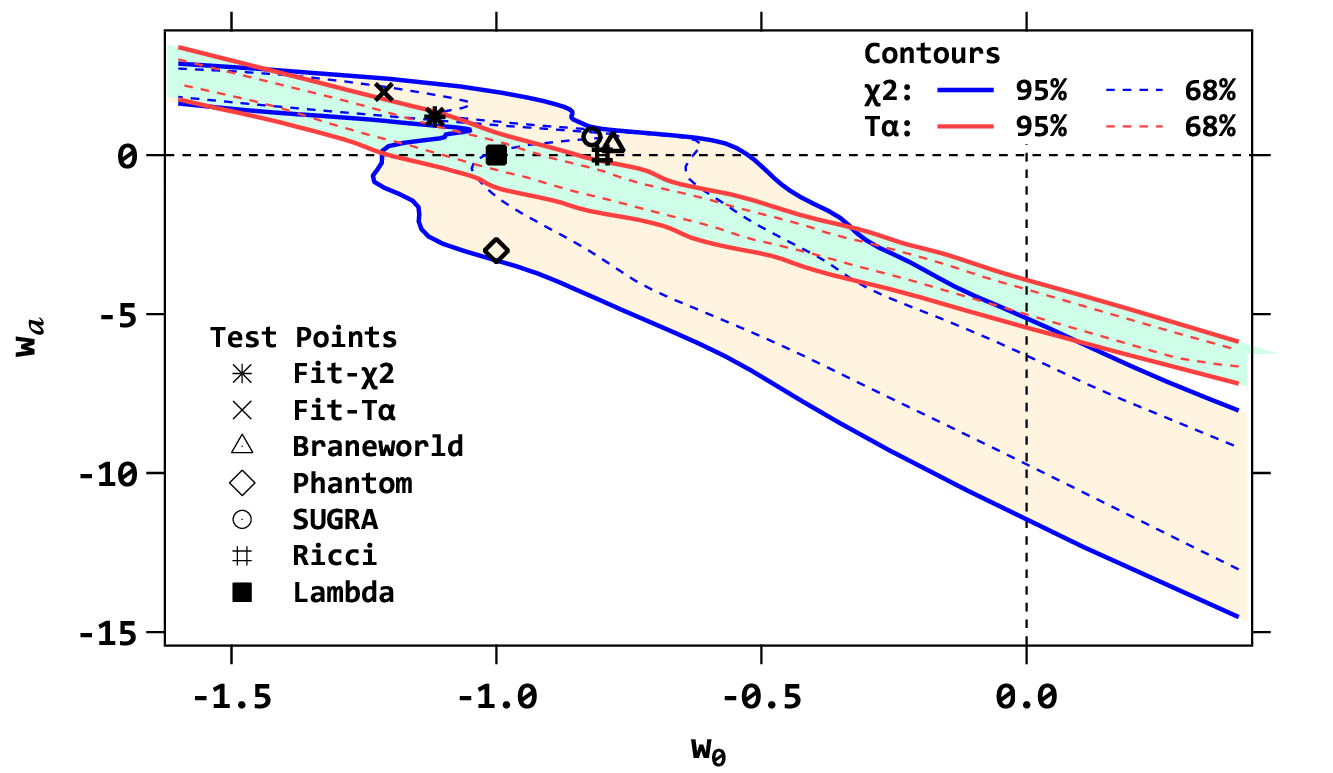}
   \caption{Monte Carlo generated 68\% and 95\% CL contours for true parameters $w_0 = -1$, $w_a = 0$  
   and points for the models mentioned in section~\ref{sec:intro}.
   }
   \label{fig:contours_null}
\end{figure}

To calculate standard deviations for comparisons against the baseline \lcdm\ model,
the Monte Carlo procedure described in section~\ref{sec:montecarlo}
was employed, where the underlying true parameters of the simulated SN catalogues were set to match those 
of \lcdm, specifically, $w_0 = -1$ and $w_a = 0$. 
Figure~\ref{fig:contours_null} displays the 68\% and 95\% CL regions
for the Monte Carlo points generated using both
the $\chi^2$ and \ta\ statistics. 
The plot includes points corresponding to the best-fit and alternative models. 

When comparing one model against an alternative model, 
it is crucial to compute the \emph{power} of the test because it quantifies 
the likelihood of detecting a signal (i.e., dynamic DE) if it exists,
which is the complement of \emph{Type II error}. 
The power of the test can be increased by reducing data errors and increasing the sample size.

The power is 
defined as $power = 1 - \beta$, where $\beta$ is the probability of 
not rejecting the null hypothesis (signal is not present) when it should have been (\emph{Type-II} error).
A dynamic DE model that predicts a small deviation from \lcdm\ (relative to
data errors) would yield low power, indicating that the model
explains the data equally as well as the baseline \lcdm.
This means that the power of the test can be used to assess
data requirements, a point which will be discussed below.

The results of the analysis are shown in 
Table~\ref{tab:stats}. The statistics in the table include the normalized
distance ($N_{\sigma}$),
the associated probability to exceed (PTE), and the power of the test.
The tests are grouped depending on the reference point used to measure the distance. 
For the first two groups, the reference point is $(-1,0)$, from \lcdm. 
For the last group, the reference point is the best-fit CPL solution using \ta, $(-1.2,2)$.
The meaning of `level of significance' for the tests in the 
second group is the significance that would be assigned to those models 
in the hypothetical situation that they were the best-fit model.
Computing this statistic is useful when evaluating 
the increase in statistical power when data errors are decreased (see section~\ref{sec:comparison_discuss}). 
\begin{table}
   \caption{Statistical tests to assess significance of alternative dynamic DE models. 
   CPL fit is the best-fit of the CPL parametrization (using the \ta\ statistic).
   $N_{\sigma}$ is the distance, in terms of number of sigmas, between the model and the 
   reference point. 
   PTE is the probability to exceed associated with the model comparison $\chi^2_p$ with $N_{dof} = 2$.
   Power is the power of the test.
   The reference point is $(-1,0)$ for level of significance tests, and $(-1.2,2)$ for compatibility with data tests.
   The Ricci geometric DE model has native parameters $0.71, -0.533$.
   }
   \label{tab:stats}
   \begin{tabular}{lccc}
   \hline
      Model  &   $N_{\sigma}$  & PTE & Power \\
   \hline
      Test: Level of significant  & & &  \\
      CPL fit     &  $2.6$  &   $0.03$               &   $0.65$  \\
   \hline
      Test: Level of significance  & & &  \\
      ~ ~ (hypothetical)  & & &  \\
      Braneworld  &  $3.1$  &   $0.009$              &   $0.79$  \\
      Phantom     &  $7.3$  &   $3 \times 10^{-12}$  &   $1$     \\
      SUGRA       &  $3.3$  &   $0.005$              &   $0.85$  \\
      Ricci       &  $2.1$  &   $0.1$                &   $0.42$  \\
   \hline
      Test: Compatibility with data  & & &  \\
      Braneworld  &  $0.9$  &  $0.7$                  &  $0.12$  \\
      Phantom     &  $12$   &  $1.33 \times 10^{-34}$ &  $1$     \\
      SUGRA       &  $1$    &  $0.6$                  &  $0.15$  \\
      Ricci       &  $0.9$  &  $0.7$                  &  $0.11$  \\
   \hline
   \end{tabular}
\end{table}

\subsubsection{Model Comparison Discussion}
\label{sec:comparison_discuss}
From the numbers in Table~\ref{tab:stats}, it can be said that the statistical significance of the
best-fit CPL model is below marginal ($2.6\sigma$). 
The entries in the table under `level of significance (hypothetical)' indicate 
that, except for the Phantom model, 
the available data
do not offer enough statistical power to discriminate against \lcdm.
The last group in the table under `Compatibility with data'
shows that while Phantom is strongly rejected, 
the data does not offer sufficient power to reject models
Braneworld, Ricci or SUGRA as incompatible with the data. 

The increased robustness and discriminatory power of the 
\ta\ statistic become apparent when one compares the PTE and power 
numbers with those obtained with the $\chi^2$ statistic. 
For example, the numbers for the Phantom model difference against the best-fit,
which show strong statistical significance using \ta, result in $N_{\sigma} = 1.7$, 
$PTE = 0.25$, and $Power = 0.3$, when using $\chi ^{2}$ effectively losing 
all statistical significance. 
Visually, this result can be 
appreciated in Figure~\ref{fig:contours_null} where the diamond symbol
for the Phantom model lies on the border of the LSQ-$\chi^2$ 95\% contour. 
In contrast, the point is well outside of the \ta\ contour. 
Similarly, for the best-fit CPL results, using LSQ-$\chi^2$, the statistical significance would 
go from marginal to negligible: $N_{\sigma} = 0.4$, $PTE = 0.9$, and $Power = 0.07$. 

The analysis has shown that with the available SN data, with
measurement errors of the order of $\sigma_m = 0.2 mag$ (average for the Pantheon+ sub-sample used in this work)
and limited catalogue depth  ($z < 2.3$), statistical tests for dynamic DE models
yield results with weak statistical power 
and marginal significance. Data requirements for more robust testing can be 
obtained by reducing the noise level ($\sigma_m$) and increasing the sample depth ($z_{max}$)
used in the Monte Carlo simulations. 
Table~\ref{tab:sigma_0p1} shows the increase in power when the 
measurement error in the simulations is reduced to $\sigma_m = 0.1$ (while keeping $z_{max} < 2.3$), an achievable target
since some of the SNe in the Pantheon+ dataset already report such a level of accuracy.
Notably, with this reduction in data errors, the statistical power to test models 
increases to a level that allows for robust and meaningful model comparison.
\begin{table}
   \caption{Statistics of model comparison against \lcdm\ for simulated data with low noise ($\sigma_m = 0.1 mag$).
   The entries in the table are as documented in Table~\ref{tab:stats}.
   }
   \label{tab:sigma_0p1}
   \begin{tabular}{lccc}
   \hline
      Model  &   $N_{\sigma}$  & PTE & Power \\
   \hline
      CPL fit     &  $6.2$   &   $3.5 \times 10^{-9}$  &   $1$  \\
      Braneworld  &  $7.6$   &   $2.3 \times 10^{-13}$ &   $1$  \\
      Phantom     &  $17.8$  &   $2.8 \times 10^{-69}$ &   $1$  \\
      SUGRA       &  $8$     &   $4.7 \times 10^{-15}$ &   $1$  \\
      Ricci       &  $5.2$   &   $1.2 \times 10^{-6}$  &   $1$  \\
   \hline
   \end{tabular}
\end{table}

\subsection{CMB Test}
\label{sec:cmbtest}
Any dynamic DE model in spatially flat geometry should predict a distance
to the last scattering surface (LSS) that is 
consistent with the same distance derived from the CMB. 
The \emph{Planck} experiment (Planck-2018) made measurements of the acoustic peaks in the CMB angular power spectrum
leading to an accurate determination of the acoustic angular scale, 
$\theta_*(z_*)$, and the comoving size of the sound horizon, $r_*$,
or the distance the photon-baryon perturbations can influence.
$z_*$ is the redshift of the LSS, which is approximately the redshift of recombination.
In flat geometry these parameters are related to the comoving radial distance to the LSS, $d_{LSS}$, as follows:
$\theta_* = r_*/d_{LSS}$. 
As shown in equations~\ref{eq:comdistance}, and~\ref{eq:integral}, the distance depends on the cosmological model and 
the value of $H_0$.
The \emph{Planck} values for these parameters are (using the TT, TE, EE+lowE+lensing result):
$z_* = 1089.92$,
$100\theta_* = 1.04110 \pm 0.00031$, and $r_* = 144.43 \pm 0.26$ Mpc,
from which a value for $d_{LSS}$ is obtained: $d_{LSS,Planck} = 13872.8 \pm 25$ Mpc.
The 1$\sigma$ uncertainty in $d_{LSS,Planck}$ was calculated using error propagation methods:
\begin{equation}
    \label{eq:d_error}
    \sigma_d = \frac{1}{\theta_*} \sqrt{\sigma^{2}_{r} + \left(\sigma_{\theta} d_{LSS}\right)^2}
\end{equation}
where $\sigma_{\theta}$ and $\sigma_r$ are the 1$\sigma$ uncertainties in $\theta_{*}$ and $r_{*}$ respectively.

To check the consistency of dynamic DE models with CMB results, 
we compute the distance $d_{LSS}$ using the best-fit CPL parameters (see Table~\ref{tab:fit}) 
$d_{LSS} = 8369 \pm 1719$ Mpc. 
The 1$\sigma$ uncertainty was estimated using a Monte Carlo procedure (based on the statistics from the Monte Carlo described in Section~\ref{sec:montecarlo}). The difference between the best fit CPL $d_{LSS}$ and $d_{LSS,Planck}$ is 5503 Mpc, which translates to a $3.2\sigma$ deviation, indicating a lack of consistency.

\section{Summary and Conclusions}
\label{sec:conclusions}
The Hubble tension is characterized by stringent constraints on both ends of the spectrum. 
On one side, the local value of $73$ \ksmpc; 
on the opposing side, the CMB-derived value of $67.4$ \ksmpc 
serves as a rigid constraint. 
One approach to addressing this discrepancy is to acknowledge the validity of both results within their respective redshift regions and seek a mechanism facilitating a transition between the two. 
We explored dynamic dark energy models as potential candidates to serve as a transitional mechanism and resolve this tension.

The CPL phenomenological parametrization of dynamic models, where \(w = w_0 + w_a(1-a)\), is simple and can represent alternative physics models. 
In the $2D$ parameter space \wa, a statistical comparison of any two models can be achieved by computing the normalized distance between points
on the \(w_0, w_a\) plane.

We introduced a statistic (the \ta\ statistic) that retains the sign of the residuals, 
enabling a reduction in the confidence region of the dark energy parameter space \wa. 
This reduction in bias errors provides more accurate information about the expansion history, 
enhancing the statistical power for model discrimination. 
Using this statistic to test the simple CPL parametrization of dynamic dark energy, 
we found a reasonable fit to Type Ia supernovae data from the Pantheon+ compilation. 
However, the data lack the necessary accuracy for a robust test, resulting in weak statistical power. 
Consequently, the best-fit CPL model is as effective as the standard \lcdm\ in explaining the data, and it falls short of alleviating the Hubble tension. 
Nevertheless, the best-fit CPL model can still play a role alongside other physical effects. 
Monte Carlo simulations demonstrate that reducing SN magnitude errors to $\sigma_{m} \sim 0.1$ mag would increase the statistical power to a level sufficient for effective model discrimination.

\bmhead{Acknowledgments}
The author thanks the hospitality and support provided by the
International Center for Relativistic Astrophysics Network in Pescara, Italy, while conducting the research for this article.

\section*{Declarations}
The authors declare that no funds, grants, or other support were received during the preparation of this manuscript.

\section*{Data Availability}
\label{sec:data}
The SNe data used in this study was obtained directly from the Pantheon github site at \\
\texttt{https://github.com/PantheonPlusSH0ES/DataRelease}

\bigskip


\bibliography{sn-bibliography}

\end{document}